\documentclass[a4paper]{article}
\usepackage{INTERSPEECH2019}
\usepackage{balance}
\usepackage{color}

\title{Single-Channel Speech Separation with Auxiliary Speaker Embeddings}
\name{Shuo Liu$^1$, Gil Keren$^1$, Bj\"orn Schuller$^{1,2}$}
\address{
    $^1$ZD.B Chair of Embedded Intelligence for Health Care and Wellbeing, \\University of Augsburg, Germany\\
    $^2$GLAM -- Group on Language, Audio \& Music, Imperial College London, UK}
\email{shuo.liu@informatik.uni-augsburg.de}

\begin{document}

\maketitle
\begin{abstract}
We present a novel source separation model to decompose a single-channel speech signal into two speech segments belonging to two different speakers. The proposed model is a neural network based on residual blocks, and uses learnt speaker embeddings created from additional clean context recordings of the two speakers as input to assist in attributing the different time-frequency bins to the two speakers. In experiments, we show that the proposed model yields good performance in the source separation task, and outperforms the state-of-the-art baselines. Specifically, separating speech from the challenging VoxCeleb dataset, the proposed model yields $4.79\,dB$ signal-to-distortion ratio, $8.44\,dB$ signal-to-artifacts ratio and $7.11\,dB$ signal-to-interference ratio.
\end{abstract}

\section{Introduction}
In the presence of two overlapping speech sources, the human brain is capable of focusing on a selected target speaker and ignoring speech from the other speaker to a large degree. 
However, constructing an automatic source separation system to extract a target speech signal from the mixture of target and interference speech signals remains a challenging task. The task becomes even more challenging when the two speakers share similar pronunciation and acoustic features. 
Conventional signal processing algorithms for this task are broadly categorised into multi-channel and single-channel methods, depending on the topology of the microphones used for the signal recording. 
Techniques based on a microphone-array such as principle component analysis (PCA), independent component analysis (ICA) and non-negative matrix factorisation (NMF) have been reported in the literature to be effective, but these techniques require some additional assumptions such as source independence, space sparsity and non-negative constrains \cite{winter2006geometrical,mitianoudis2004audio,weninger2012optimization,ozerov2010multichannel}.
On the other hand, traditional single-channel source separation approaches such as computational auditory scene analysis (CASA) exploits pitch and onset as grouping cues to decompose a mixed speech signal \cite{hu2010tandem}.  

Over the past few years, neural network models have managed to outperform classic signal processing algorithms in many speech related applications, such as automatic speech recognition (ASR) \cite{zhang2017towards,seltzer2013investigation}, speech enhancement \cite{Keren2018,araki2015,weninger2015speech}, and speech emotion recognition (SER) \cite{zhang2018deep,han2017reconstruction,tzirakis2017end}.
This recent success of neural network models has inspired some of the recent solutions for the single-channel source separation task \cite{Wang2018SupervisedSS,lee2017fully,samui2017deep}.
A prominent recent work introduces the deep clustering approach \cite{hershey2016deep}. In this work, an assumption is made that each time-frequency bin in the speech spectrogram belongs to only one speaker. A deep recurrent network is trained to produce an embedding vector for each time-frequency bin (TF-bin), and the model is trained to emit similar embeddings for TF-bins that originate from the same speaker. A k-means clustering algorithm is then applied at test time to attribute the TF-bins to the different speakers according to the learnt embeddings.
The deep attractor network (DaNet) \cite{chen2017deep} extends the deep clustering framework. In this work, embeddings for the different TF-bins that originate from the same speaker are averaged at training time to create an attractor for this speaker. The model is trained to produce embeddings for the TF-bins that are similar to their respective attractors. The DaNet was shown in experiments to yield superior source separation performance compared to the deep clustering method. 

\begin{figure*}[h!]
  \centering
  \includegraphics[width=\linewidth]{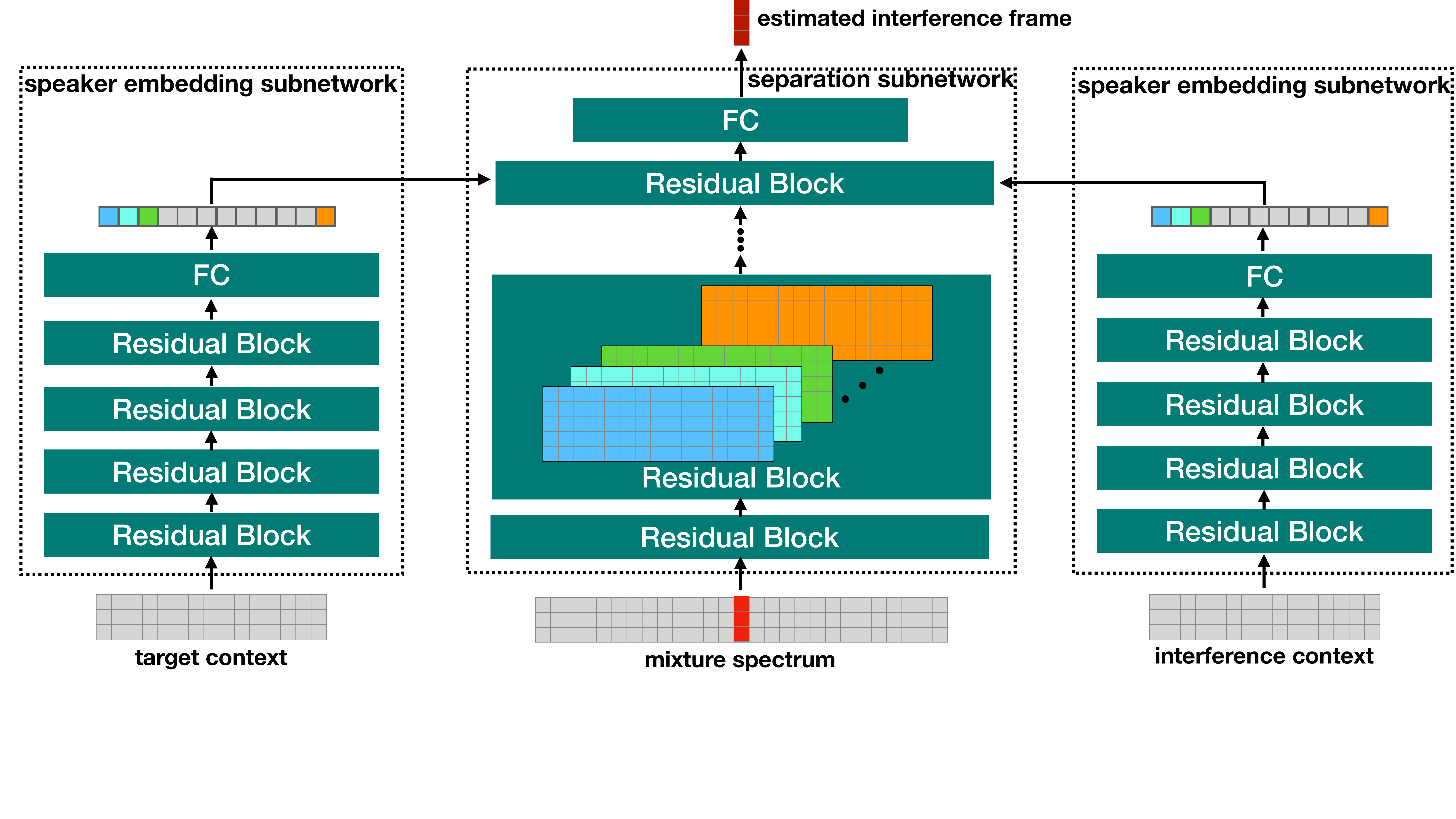}
  \vspace{-2cm}
  \caption{The source separation model architecture. The identical speaker embedding subnetworks processes the target and interference contexts via a sequence of $4$ residual blocks to produce target and interference speaker embeddings. The separation subnetwork processes the mixture segment through a sequence of $8$ residual blocks, each additionally conditioned on the target and interference speaker embeddings, to output an estimated interference frame. The difference between the central frame of the mixture segment and the estimated interference frame is the estimated target frame.} 
  \vspace{-1em}
  \label{fig:model}
\end{figure*}

However, both the deep clustering and DaNet methods are facing a non-trivial challenge. Both models have to learn to produce similar embeddings to TF-bins that belong to similar speakers, without any indication in what region in the embedding space the similar embeddings should be. As embeddings for TF-bins that belong to different speakers may concentrate in different regions of the embedding space, given a new speaker at test time, different regions in the audio segment will have to agree in which area of the embedding space the embeddings for this speaker should concentrate. In order for this to happen, bottom layers of the neural network will have to exchange information between all the different parts of the audio segment, communicating the target area in the embedding space. As the network is not explicitly guided to do so, learning this behaviour from data alone is indeed a non-trivial challenge. 

In this paper, we propose a novel source separation model based on stacks of residual blocks.
To alleviate the above issue, we supply the model with short additional context recordings containing clean speech from each of the speakers. Each of the context recordings is then processed with a dedicated subnetwork to create a speaker embedding vector. The main subnetwork processes the mixed speech segment and emits the separated speech, and is also conditioned on the speaker embedding vectors, that assist in attributing the different TF-bins to the different speakers, thus circumventing the issue described above. It is important to note that supplying the network with those additional inputs is a realistic setting. Indeed, in many real-life applications the audio segments for separation contain large parts where only a single speaker is present. Those parts can be extracted and used by the model as the additional speaker context. In experiments, we find that our proposed model outperforms the state-of-the-art deep clustering and DaNet models in the source separation task using a large-scale speech dataset. 

The rest of the paper is organised as follows. Section \ref{sec:data} presents the data used in this work and its processing. Section \ref{sec:model} discusses the structure of the proposed model, experiments and evaluation for the source separation task are described in Section \ref{sec:experiments}. Final conclusions are drawn in Section \ref{sec:conclusions}.

\section{Data description and processing} \label{sec:data}
In general, the performance of a deep neural network model for source separation improves as the size and diversity of the speech data increases. The VoxCeleb dataset \cite{DBLP:conf/interspeech/NagraniCZ17, DBLP:conf/interspeech/ChungNZ18} provides more than 2000 hours of single-channel recordings extracted from Youtube interviews of more than 7000 speakers, and includes more than one million utterances.
The dataset contains two versions, VoxCeleb1 and VoxCeleb2, each with its own training and test set. As each version consists of distinct speakers, our training and test sets are comprised of the union of the two corresponding sets in the two versions. For the creation of the validation set, 20 utterances from different speakers were chosen from the training set. The speakers with utterances appearing in the validation set were removed from the training set. The above procedure ensures that the training, validation and test sets are speaker independent. The amount of data for validation is selected to be very small because we do not require much hyper-parameters tuning for our proposed source separation network. The utterances in VoxCeleb cover different nationalities and range from 4 to 12 seconds in length. 

At training time, the model inputs are created using a random process at each iteration, for improving the separation quality by creating a larger effective size for the training set. 
At each iteration, we randomly sample two speakers, named the \emph{target speaker} and the \emph{interference speaker}, and a random utterance from each of the two speakers, named the \emph{target utterance} and the \emph{interference utterance} respectively. 
The two utterances are truncated to have the same length, and then the interference utterance is mixed into the target utterance using a random SNR of either -5 ,0, 5, 10, 15 or 25\,dB to create the \emph{mixture utterance}. 
The log magnitude spectrum is extracted from each of the target, interference and mixture utterances by applying a short-time Fourier transform (STFT) on each segment, using a 25\,ms Hanning window shifted by 10\,ms. Prior to computing the logarithm, a small value of $10^{-5}$ is added to the magnitude of the STFT output, to prevent the model from fitting to imperceptible differences in magnitude. The sampling frequency of all audio segments is 16\,kHz, therefore each frame consists of 400 samples and its resulting feature vector is comprised of 201 frequencies. 

The source separation model processes three inputs. The \emph{mixture segment} consists of 100 successive frames from the mixture spectrum, chosen randomly at each iteration from the mixture utterance. 
The parts of the target and interference utterance that were used to create the mixture segment are named the \emph{target segment} and the \emph{interference segment} respectively. 
The \emph{target context} is a 35 frames segment chosen randomly at each iteration from the part of target utterance that does not appear in the mixture segment. The \emph{interference context} is chosen from the interference utterance using the same process. The target and the interference contexts each contain speech from a single speaker, and are used by the model to create the speaker embeddings to assist with the source separation process, as described in Section \ref{sec:model}. It is important to note that in many real-life applications the audio segment for separation contains large parts where only a single speaker is present, therefore conditioning our model on the target and interference contexts, that contain speech from a single speaker each, is a realistic setting that may be deployed in a variety of real-life applications. 
The target segment is used by the separation model as the training label. 

For creating the validation and test sets, all utterances from the corresponding set were randomly split into pairs of target and interference utterances, such that each pair contains utterances from different speakers. Each pair was then mixed to create the mixture segment, using an SNR of either -5, -3, -1, 0, 1, 3$ or $5\,dB. These SNR values ensure a fair comparison of our algorithm with previous work. A larger variety of SNRs were used in the training process, to facilitate the trained model's robustness to a variety of SNRs. Target and interference contexts were chosen to be from the beginning of the target and interference utterances, respectively. Test and validation sets were created once, and were consistent across all experiments.

\section{Source separation model} \label{sec:model}
Residual neural networks (resnets) introduce shortcut connections to the conventional CNN framework and enable a substantially deeper architecture, which has been validated to be successful in both the computer vision and audio domains\cite{he2016deep,vydana2017residual,jung2017resnet}.
A basic residual block contains two convolutional layers, where batch normalisation \cite{ioffe2015batch} followed by a rectified linear unit (ReLU) are applied between the convolutional layers.  The residual block's input is added to the output of the second convolutional layer, and again batch normalisation and ReLU activation are applied to emit the block's output. In this work, we use two-dimensional convolutional layers that operate on the time and frequency axes. 

The architecture of our proposed source separation model is based on stacks of residual blocks as depicted in Figure \ref{fig:model}. The model consists of three subnetworks, each mainly contains a sequence of residual blocks. First, a speaker embedding subnetwork processes the target context, to emit the \emph{target speaker embedding}. Similarly, an identical subnetwork processes the interference context to emit the \emph{interference speaker embedding}. The separation subnetwork then processes the two speaker embeddings and the mixture segment to emit an \emph{estimated interference frame}. The estimated interference frame is an estimate of the interference components in the central frame of the mixture segment. Finally, the \emph{estimated target frame} is computed to be the difference between the central frame of the mixture segment and the estimated interference frame. The model is trained to minimise the squared difference between the estimated target frame and the true target frame that is the central frame of the target segment. 

\subsection{Speaker embedding subnetwork}
Each of the two speaker embedding subnetworks takes a speech context as input, and its output is a speaker embedding vector that may contain valuable acoustic information obtained from the speech segment.
The speaker embedding subnetwork is compromised of a sequence of four residual blocks using the specifications from Table \ref{tab:embedding}.
The output feature maps of the last residual block is averaged across all locations (time steps and frequency bins) to get a speaker embedding vector with a fixed length of 512.

\begin{table}[th]
  \caption{The speaker embedding subnetwork specifications. The subnetwork contains 4 residual blocks, each with different kernel size, stride, and number of channels.}
  \label{tab:embedding}
  \centering
  \begin{tabular}{ l c c c}
    \toprule
    \multicolumn{1}{c}{\textbf{Block}} & 
                                         \multicolumn{1}{c}{\textbf{Kernel}} & \multicolumn{1}{c}{\textbf{Stride}} & \multicolumn{1}{c}{\textbf{\#Channels}} \\
    \midrule
    $1$ & $(8,4)$ & $(3,2)$ & $64$\\
    $2$ & $(8,4)$ & $(3,2)$ & $128$\\
    $3$ & $(4,4)$ & $(1,1)$ & $256$\\
    $4$ & $(4,4)$ & $(1,2)$ & $512$\\
    \bottomrule
  \end{tabular}
\end{table}

Two identical such subnetworks (each with its own learnable parameters) process the the target and interference contexts and produce the target and interference speaker embeddings. The learnt target and interference speaker embeddings are injected into the separation subnetwork to assist with the source separation task.

\subsection{Separation subnetwork} 
This separation subnetwork is comprised of a sequence of 8 residual blocks that process the mixture segment, using the specifications from Table \ref{tab:separation}. 
For each convolutional layer, the learnt target and interference speaker embeddings are each linearly projected to a dimension equal to the number of feature maps in this layer by applying a trainable fully-connected layer. The two projected vectors are then added to every location in the output map of the convolutional layer. 
By doing so, all convolutional layers in the separation subnetwork are conditioned on the information of target and interference speech, allowing the separation network to better estimate the components belonging to the interference speaker.

\begin{table}[th]
  \caption{The separation subnetwork specifications. The subnetwork contains 8 residual blocks, each each with different kernel size, stride and number of channels.}
  \label{tab:separation}
  \centering
  \begin{tabular}{ l c c c}
    \toprule
    \multicolumn{1}{c}{\textbf{Block}} & 
                                         \multicolumn{1}{c}{\textbf{Kernel}} & \multicolumn{1}{c}{\textbf{Stride}} & \multicolumn{1}{c}{\textbf{\#Channels}} \\
    \midrule
    $1$ & $(4,4)$ & $(1,1)$ & $64$\\
    $2$ & $(4,4)$ & $(1,1)$ & $64$\\
    $3$ & $(4,4)$ & $(2,2)$ & $128$\\
    $4$ & $(4,4)$ & $(1,1)$ & $128$\\
    $5$ & $(3,3)$ & $(2,2)$ & $256$\\
    $6$ & $(3,3)$ & $(1,1)$ & $256$\\
    $7$ & $(3,3)$ & $(2,2)$ & $512$\\
    $8$ & $(3,3)$ & $(1,1)$ & $512$\\
    \bottomrule
  \end{tabular}
\end{table}

At last, we flatten the output of the last residual block and feed it through a fully-connected layer to get a 201 dimensional output. This output is added to the central frame of the input mixture segment to yield the estimated target frame. During the training phase, we optimise the network parameters to minimise the mean squared error (MSE) between the estimated target frame and the central frame of the target segment, using stochastic gradient descent (SGD) with a learning rate of $0.1$. 
At validation and test time, inverse Short-Time Fourier transform (iSTFT) was used to reconstruct the target speech using the phase of the mixture segment. 

\section{Experiments and results} \label{sec:experiments}

\begin{figure*}[h!]
  \centering
  \hspace*{-0.2cm}
  \includegraphics[scale=0.36]{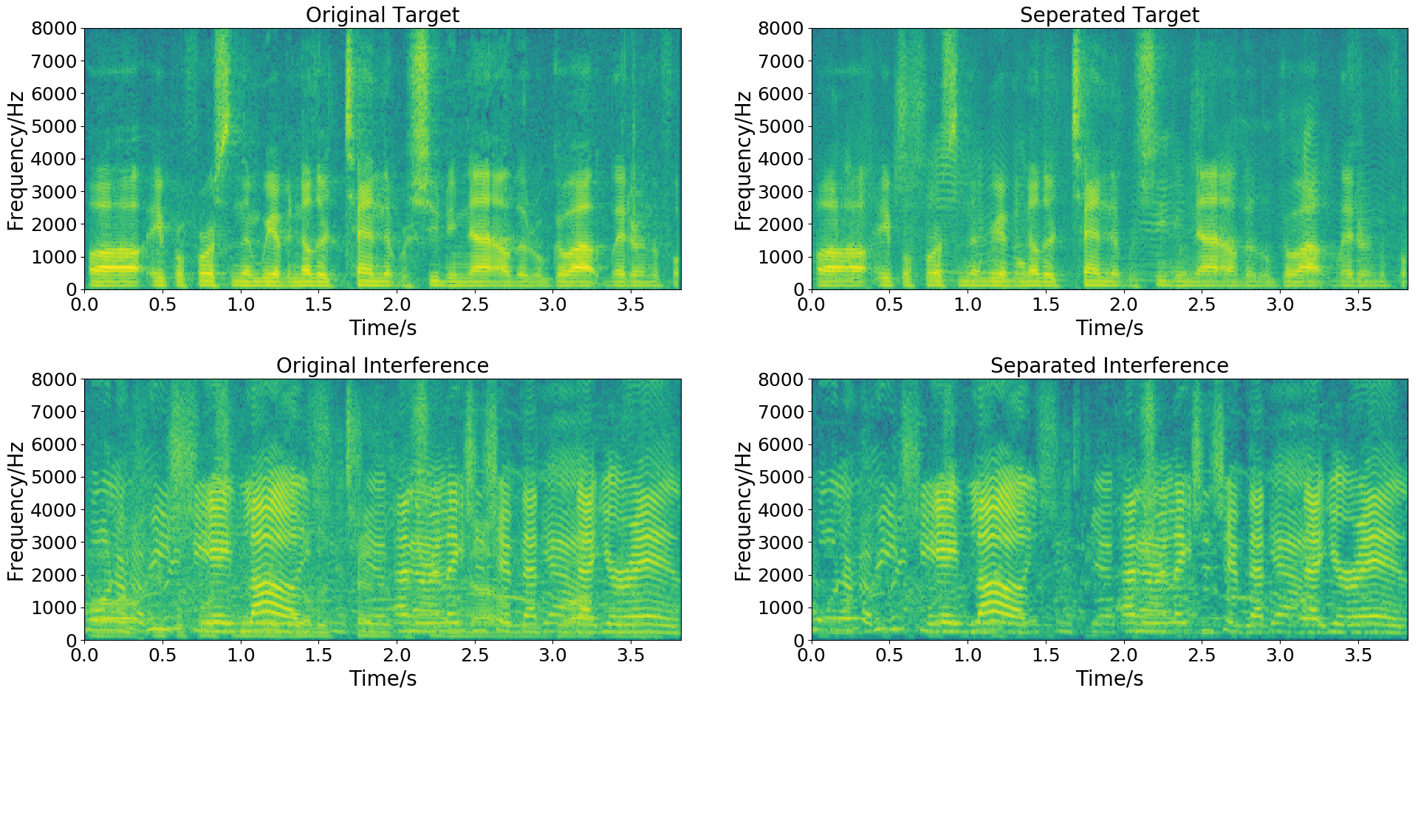}
  \vspace{-2.3cm}
  \caption{An example of true target and interference spectrum, and the corresponding output of the source separation model.}
  \vspace{-0.8em}
  \label{fig:performance}
\end{figure*}

We conduct experiments for evaluating the effectiveness of the proposed source separation model. The performance of both our proposed model and the state-of-the-art baselines recently proposed for source separation \cite{hershey2016deep,chen2017deep} are compared in a large-scale source separation task using the VoxCeleb dataset and unseen speakers at test time as was described in Section \ref{sec:data}.

\subsection{Baseline models}
As a baseline, we implemented the deep clustering model and the deep attractor network (DaNet) as described in \cite{hershey2016deep,chen2017deep}. For experiments with the two baseline methods, we follow the audio preprocessing pipeline from these works. All audio segments were downsampled to 8\,kHz before processing and then mixed using one of the SNRs -5, -3, -1, 0, 1, 3, 5\,dB. Log magnitude spectrum of the mixture speech was obtained using STFT with a 32\,ms window length with an 8\,ms window shift. Chunks of 100 frames were used as input to the model. 
The deep clustering model uses 2 bidirectional LSTM layers, while DaNet contains 4 bidirectional LSTM layers, both with 600 hidden units in each layer, to learn 20-dimensional embeddings for every TF-bin of the mixture spectrum. The deep clustering model is trained to produce similar embeddings to TF-bins that originate from the same speaker. The DaNet model averages the embeddings that originate from similar speakers to create attractors, and the model is trained to produce  embeddings that are similar to their appropriate attractors. Optimisation parameters were chosen using the validation set. At test time, k-means clustering is used to allocate the TF-bins to the two speakers based on their embedding vectors. 

\subsection{Comparison}
We consider three objective evaluation metrics for separation as described in \cite{vincent2006performance}: signal-to-distortion ratio (SDR), signal-to-artifacts ratio (SAR) and signal-to-interference ratio (SIR). 
We use the BSSEval toolbox \cite{stoter20182018} to compute the three evaluation metrics for the VoxCeleb test set and each of the evaluated models. 
Table \ref{tab:compare} contains the results for all the evaluated models. 
As seen in the results table, our proposed model outperforms the deep clustering and DaNet baselines by a large margin in terms of SDR and SAR, which measures the lack of distortion and algorithm artifacts in the recovered speech signals, concerning speech quality. Regarding SIR, which represents the ability of the model to suppress the interference speech, the DaNet model yielded a better value compared to our approach. However, SIR does not measure whether the target speech was preserved, therefore a high SIR value with a low SAR and SDR value indicates that the DaNet model suppresses both the interference and the target speakers to a large extent. We note that both baseline approaches were reported to yield good source separation performance in previous work. However, in this work we use a challenging dataset of natural speech, which may be the reason for the relatively low performance of these methods in this work. 


\begin{table}[th]
  \caption{Comparison of test set evaluation metrics (in dB). DC and DaNet stand for deep clustering and deep attractor network. SDR, SAR, and SIR represent signal-to-distortion, signal-to-artifacts, and signal-to-interference ratio, respectively.}
  \label{tab:compare}
  \centering
  \begin{tabular}{ l r r r }
    \toprule
    \multicolumn{1}{c}{\textbf{   }} & \multicolumn{1}{c}{\textbf{SDR}} & \multicolumn{1}{c}{\textbf{SAR}} & \multicolumn{1}{c}{\textbf{SIR}}\\
    \midrule
    \textbf{DC\cite{hershey2016deep}}    & 0.84 & 2.09 & 6.58 \\
    \textbf{DaNet\cite{chen2017deep}}    & 1.81 & 3.29 & \textbf{10.41} \\
    \textbf{Proposed} & \textbf{4.79} & \textbf{8.44} & 7.11 \\
    \bottomrule
  \end{tabular}
\end{table}



To better visualise the performance of our source separation algorithm, we depict an example of target and interference segments from two speakers in the test set on the left column of Figure \ref{fig:performance}, where the target speaker is male and the interference speaker is female. 
The right column shows our model's output, the estimated (separated by the model) target and interference segments. 

\section{Conclusions} \label{sec:conclusions}
In this paper, we developed a single-channel source separation model that uses additional conditioning on separate speaker context recordings. 
The model learns to create speaker embeddings for unseen speakers from additional context recordings. The speaker embeddings may contain important acoustic information regarding the different speakers, and assist the model in enhancing the source separation quality.
In experiments, our proposed model considerably outperforms two state-of-the-art baseline models. 
Future work should focus on extending this method to operate with any number of speakers, and further improving separation quality on the challenging dataset that was used. Moreover, more work should be done on maintaining speech intelligibility in the source separating process, to overcome distortions that are occasionally introduced in the audio. Furthermore, the speaker embeddings may be further investigated, for better understanding how the model codes the different speakers, and what specific acoustic information assists the model with the separation process.


\balance
\bibliographystyle{IEEEtran}

\bibliography{mybib}

\end{document}